\documentstyle[epsfig,longtable]{aipproc}

\begin{document}

\title{Dual Accretion Disks in Alternate Gravity Theories}
 
\author{James S. Graber$^*$}
\address{$^*$407 Seward Square SE, Washington, DC 20003}

\maketitle

\begin{abstract}
The interior of gravitationally collapsed objects in 
alternate theories of gravity in which event horizons 
and singularities do not occur in strong field gravity
were generically investigated.  These objects, called 
red holes, were found to contain dynamic configurations 
of matter, radiation and spacetime similar to inside out 
accretion disks well inside the photon orbit.  Applications
to astrophysical phenomena are briefly described.
\end{abstract}

\section*{NATURE MIGHT PREFER RED HOLES
}

Rather than general relativity, an alternate theory of gravity -- 
in which event horizons do not form around collapsing compact 
bodies -- might be true.  In this case, black holes do 
not exist in nature.  Instead, compact bodies that collapse 
gravitationally inside their photon orbit form a different 
configuration of matter, radiation and space-time, which can be 
called a red hole.  The literature contains several theories, 
consistent with all known experimental evidence, in which red 
holes occur instead of black holes\cite{yil,ros,mof}.

\section*{THREE PARADIGMS FOR GRAVITATIONAL COLLAPSE}
The black hole paradigm for gravitational collapse is now well 
known.  In the black hole paradigm, once a certain critical 
density is reached, a compact body collapses forever, reaching 
infinite redshift in a finite time and forming an event horizon, 
which is a one-way sink for matter, energy and radiation -- a sort 
of bottomless pit.  This is the most popular expectation today.  

In the frozen star paradigm, which preceded the black hole 
paradigm, gravitational collapse ended with a super-hard kernel, 
sort of like a tougher, denser, more massive neutron star.  
Variants of this paradigm are still considered today.  See, for 
example, Robertson\cite{rob}.  

In the less well known red hole paradigm, gravitational collapse 
results in a configuration of matter, energy and space-time 
that contains no solid surface, no singularities, and no 
event horizons. Unlike the frozen star paradigm, red holes 
do not stop collapsing and form a dense kernel that resembles 
a denser, tougher neutron star. Unlike the black hole paradigm,
 red holes do not collapse 
forever and form a bottomless pit with an event horizon as a 
one-way sink for matter, energy and radiation.  

Instead, the matter and radiation inside a red hole form a 
dynamical configuration somewhat resembling a globular cluster or an acretion flow.  
Orbits inside a red hole are typically highly elliptical, deeply 
plunging, roseate shaped and at relativistic velocities.  
Space-time is highly stretched and distorted in a manner very similar to 
the parts of a black hole that are inside the photon orbit but 
outside the event horizon.  As a result of this progressive space-time stretching, the density of a red hole decreases toward the 
center, and the center of a red hole can be -- and usually is -- a 
near-vacuum.

\section*{WHEN WILL A RED HOLE STOP COLLAPSING?}

As a star collapses into a red hole, its locally apparent density 
increases until it crosses the photon orbit and then it starts to 
decrease.  It reaches a density much greater than a neutron 
star's, and then the density decreases again to that of a neutron 
star, then a white dwarf, then a normal star, and eventually to 
the density of a collisionless plasma.  Once the density drops to  
a level appropriate to a collisional gas or plasma, a violent 
relaxation will take place, and the matter and radiation deep 
inside the red hole will assume a randomized near-equilibrium 
configuration similar to that of a globular cluster (an inside-out 
globular cluster, as the matter is denser on the outside than on 
the inside).  Self-organizing criticality will keep the 
equilibrium near the point of marginal collisionality.  Below the 
collisionality point, the matter will scatter to a more tightly 
bound configuration only very slowly.  Above collisionality, 
scattering will rapidly bifurcate the red hole contents into a 
more energetic escaping fraction and a less energetic, less dense, 
more tightly bound remainder, which will thereafter evaporate only 
very slowly.

\section*{WHY DOES A RED HOLE STOP COLLAPSING?}

In red hole theories, only infinite densities lead to infinite 
redshift and hence the possibility of an event horizon or a 
bottomless pit or an endless collapse.  In red hole, theories once 
gravitational forces have overcome all other forces and collapsed 
inside the photon orbit occurs, the space-stretching effects 
happen faster than the infall rate.  Contrary to intuition, the 
density then begins to decrease rather than increase.  This allows 
the collapse to eventually stop when the density is once again low 
enough to allow the particles to follow essentially independent 
trajectories.

In principle, there is no limit to how near a configuration can 
approach infinite density and infinite redshift. In fact, a 
hypothetical point particle will itself have infinite density and 
hence an infinite redshift and a point singularity.  Thus, one 
may have to appeal to quantum uncertainty principle 
limits or string theory dualities to make singularities, infinite 
densities and infinite redshifts impossible.  

In practice, however, if one begins with real electrons, baryons 
and photons in reasonably random thermal motion, no infinite 
density state will occur, even in gravitational collapse.  The 
center of a red hole will have a nearly flat potential like the center of a 
spherical shell, and kinematics much like an 
inside-out globular cluster.  Therefore, the density and hence the
redshift and also the stoppage of the collapse will be 
determined by an equilibrium near the point of marginal 
collisionality.

\section*{WHAT DOES A RED HOLE LOOK LIKE?}

From the outside, a red hole looks like a cross between a neutron 
star and a black hole: Like a black hole, a red hole has no solid 
surface, 
and a red hole is entirely within its photon orbit, as well as its 
innermost stable circular orbit.  Like a neutron star, 
a red hole emits some of the matter and radiation that fall into it,
 and 
a red hole may be optically thick and have a visible last scattering
 surface.  
Mathematically, a red hole looks like a black hole with a cutoff. 
 Think of the Thorne, et al., membrane approximation to a general 
relativity black hole\cite{th}.  Put a hollow sphere around the black hole, 
just outside the event horizon, and connect opposite points with a 
straight line.  This is a good model of a red hole.  

To understand what happens inside a red hole, think particularly 
of the loss cone model of Misner, Thorne and Wheeler\cite{mtw}.  Only matter 
moving at near-relativistic speeds can escape the red hole.  Only 
matter and radiation directed near vertically can escape.  The 
size of the loss cone decreases rapidly inside the photon orbit.  
Hence, only a small fraction of matter and radiation can escape.  
(Think of it as similar to Hawking evaporation.)  Most matter and 
radiation are trapped inside the red hole for a long time.  

\section*{THE INNER "ACCRETION DISK"}

The inside of a red hole looks a lot like an inside-out accretion 
disk with a near-vacuum at the center.  Since there is no singularity or event horizon to swallow up the 
matter and radiation falling into a red hole, it can only escape 
by reemerging from the red hole, of which there is only a very 
small probability.  Since the gravitational force inside the photon orbit is too high 
for anything to remain static, all the contents are in motion, 
mostly at relativistic speeds.  A nonrotating red hole might 
resemble a globular cluster or a spherical galaxy inside a deep 
potential well.  A rotating red hole might resemble an accretion 
disk or a spiral galaxy similarly situated.  

However, there are no circular orbits inside a red hole, only 
deeply plunging roseate shaped orbits.  Also contrary to 
intuition, a red hole does not get denser and denser as the center 
is approached.  As the center is approached, the space-bending and 
space-stretching effects increase faster than the rate of approach 
to the center, thus causing the density to decrease.  Thus the 
center of a red hole is hollow -- in fact resembling a near-vacuum 
-- and the slope of the gravitational potential flattens, as is 
expected inside a spherical shell.  This is why no singularity 
forms and why the redshift at the center stops at a finite value 
and does not approach infinity.  Hence the matter inside a red 
hole near equilibrium is densest on the outside and least dense on 
the inside, which is why it resembles an inside-out accretion 
disk.

\section*{RED HOLES WORK BETTER THAN BLACK HOLES}

Red holes can explain many astrophysical phenomena better than 
black holes of similar mass.  Because more energy can escape from 
a red hole, there is more energy to power supernovae, jets, and 
gamma ray bursts.  Because you can see inside a red hole, small 
size scale and rapid variability effects can be explained.  
Because accretion disks can have an active center -- instead of a 
solid body or a bottomless pit -- it is easier to form, power, and 
collimate jets.

If two particles of equal mass originate near the boundary of the 
red hole (i.e. just marginally bound or just marginally unbound), 
fall in, reach relativistic velocities, and scatter so that one 
particle is tightly bound and the other one escapes, the escaping 
particle will have excess kinetic energy of the same order of 
magnitude as its rest mass.  Likewise if the two particles scatter 
inelastically and emit a photon that can escape, its energy upon 
emerging will be of the order of magnitude of the rest mass of the 
particles or of the binding energy at the level to which they have 
fallen when they scatter.

Hence it is possible that if roughly half the matter in an initial 
collapse is trapped in the red hole and half is scattered out, 
about one quarter to one half of the rest mass of the collapsing 
matter will be converted into escaping kinetic energy and 
radiation available to drive supernovae explosions, gamma ray 
burst fireballs, jets, superluminal ejected blobs, or other 
energetic phenomena.

\end{document}